\documentclass[a4paper]{article}

\usepackage{INTERSPEECH2018}

\title{Multiple Instance Deep Learning for Weakly Supervised Small-Footprint Audio Event Detection}

\name{ 
    Shao-Yen Tseng$^{\dag}$,
    Juncheng Li$^{\ddag\gamma}$,
    Yun Wang$^{\gamma}$,    
    Florian Metze$^{\gamma}$,
    Joseph Szurley$^{\ddag}$,
    Samarjit Das$^{\ddag}$
}
\address{
$^{\ddag}$Robert Bosch LLC, Research and Technology Center, USA  \\
$^{\dag}$University of Southern California, Department of Electrical Engineering, USA \\
$^{\gamma}$Carnegie Mellon University, Language Technology Institute, USA
}
\email{shaoyent@usc.edu, \{junchenl,yunwang,fmetze\}@cs.cmu.edu, \{joseph.szurley,samarjit.das\}@us.bosch.com}

\begin{document}

\maketitle
\begin{abstract}
State-of-the-art audio event detection (AED) systems rely on supervised learning using strongly labeled data. 
However, this dependence severely limits scalability to large-scale datasets where fine resolution annotations are too expensive to obtain.
In this paper, we propose a small-footprint multiple instance learning (MIL) framework for multi-class AED using weakly annotated labels.
The proposed MIL framework uses audio embeddings extracted from a pre-trained convolutional neural network as input features.
We show that by using audio embeddings the MIL framework can be implemented using a simple DNN with performance comparable to recurrent neural networks.

We evaluate our approach by training an audio tagging system using a subset of AudioSet, which is a large collection of weakly labeled YouTube video excerpts.
Combined with a late-fusion approach, we improve the F1 score of a baseline audio tagging system by 17\%.
We show that audio embeddings extracted by the convolutional neural networks significantly boost the performance of all MIL models. 
This framework reduces the model complexity of the AED system and is suitable for applications where computational resources are limited.

\end{abstract}

\section{Introduction}
\label{sec:intro}
Increasingly, devices in various settings are equipped with auditory perception capabilities.
The inclusion of acoustic signals as an extra modality brings robustness to a system and offers improved performance in many tasks. 
This benefit can be attributed to the omnidirectional nature of acoustic signals which provides a valuable cue for detecting events in various applications.
For example, \cite{Ubhayaratne_AudioSignal} analyzed audio signals to monitor the conditions of industrial tools, and in \cite{SEYOUM2017455} a water leakage detection system using sound recordings of water pipes was proposed. 
Such systems are able to run in real-time and at a lower cost as capturing audio is much less expensive than distributing specialized physical sensors throughout the environment.
In addition, acoustic signals can provide informational cues that are hard to or cannot be captured by other modalities. 
A common example is the detection of alarms or sirens in a driving scenario with smart cars.
Very often, sources of these warning sounds may be visually occluded and these events are only detectable using auditory perception \cite{Meucci_A-real-time-siren-detector}\cite{Schroder_Auto-acoustic-siren-detection}.
Many of these applications also have a requirement of real-time operation using low computational resources. 
This is a major challenge since, unlike human speech, environmental sounds are much more diverse and span a wider range of frequencies.
Audio events that occur in these settings are also usually sporadic and corrupted by noise. 

Previous works on AED have relied on training models using a supervised learning paradigm which requires strongly labeled data \cite{portelo2009non}\cite{dikmen2013sound}. 
However, given the difficulty and high resource requirement of annotating large datasets there are only a few datasets that are publicly available and are often of limited size \cite{salamon2014dataset}\cite{mesaros2016tut}. 
Motivated by this, many recent works have explored the use of weakly labeled data for training AED systems.
One successful approach is to transform the audio into time-frequency representations and apply a convolutional recurrent neural network to tag or classify the entire clip \cite{7933050}\cite{Choi_CRNN}.
These methods, however, are unsuitable for real-time applications as the recurrent and subsequent pooling layers require the full clip to be parsed before a decision can be made. 
In addition, the complexity and computation time of these models are quite high.
Another approach for learning with weak labels is to treat segments in an audio clip as a \textit{bag of instances} and apply multiple instance learning \cite{babenko2008multiple}. 
The MIL model assumes independent labels for each instance and accounts for the uncertainty of the weak labels by assigning a positive bag label only if there is \textit{at least one} positive instance.
Evidently, this paradigm is more suitable for portable applications as the classifier can be applied to individual instances which is ideal for real-time operation.

In this work, we propose to enhance the framework for multi-class MIL using convolutional audio embeddings.
Different from prior works, our proposed architecture addresses the issue of building low complexity models with a small footprint for real-time applications. 
We propose the use of audio embeddings as input features and show that by using pre-trained embeddings the MIL model can be implemented with a simple DNN architecture. 
The use of audio embeddings also significantly improves AED accuracy compared to random initialization. 
Our proposed architecture removes the need for complex CNN structures or recurrent layers which drastically reduces model complexity and is suitable for portable applications with low computational resource and real-time requirements.





\section{Multiple Instance Learning}
\label{sec:mil}

\subsection{MIL Framework}
The task of detecting audio events using weakly labeled training data can be formulated as a multiple instance learning problem \cite{raj_audio-event-detection}.
In MIL, labels are assigned to \textit{bags} of \textit{instances} without explicitly specifying the relevance of the label to individual \textit{instances}. 
All that is known is one or more \textit{instances} within the \textit{bag} contribute to the \textit{bag} label.
Applying this framework to our task, we view audio clip $i$ as a \textit{bag} of 
\textit{instances} $B_i = \{x_{ij}\}$ where each \textit{instance} $x_{ij}$ is an audio segment $j$ of shorter duration. 
We then assign all the labels of the clip to the bag so that each bag has the label $Y_i = \{y_{in}\}$ where $y_{in} = 1$ indicates the presence of audio event $n$.
The goal of the MIL problem is then to classify labels of unseen \textit{bags} given only the \textit{bag} and label pairs $(B_i,Y_i)$ as training data.
In this we work we implement the MIL framework using neural networks. 

\subsection{MIL using Neural Networks}

In our implementation we generate instances by segmenting the audio clip into non-overlapping 1-second segments and taking the time-frequency representations.
The segment size was chosen as a balance between number of total instances and coverage of audio events. 
We use a frame size of 25ms with 10ms shift in the short-time Fourier transform and integrate the power spectrogram into 64 mel-spaced frequency bins.
A log-transform is then applied to the spectrogram.
We also use the first delta as an additional input channel. 

Since the spectrogram can be viewed as an image we employ convolutional layers for feature extraction.
We reference CNN architectures proven to have good performance in the field of computer vision.
Specifically, we use the first three conv groups from VGG-16 \cite{Simonyan2014Very-Deep-Convo} and add two fully-connected layers of size 3072 and 1024.
Batch normalization is added after each convolutional layer.
The ReLU activation function is used in all layers. 
As our goal is a multi-label system we apply a sigmoid activation function and view the outputs as independent posterior probability estimates for each class. 
We use a reduced version of the full VGG model because (1) we are exploring compact models for portable applications and (2) the subset dataset does not contain enough samples to train large models without overfitting. 

To obtain a prediction for the entire bag we adopt a na\"ive approach and assign the label of the maximum scoring instance to the bag. 
The motivation behind this is in part due to the fact that since instances in a continuous audio clip are not i.i.d. many MIL algorithms are not applicable \cite{babenko2008multiple}.  
However this approach is still beneficial as it allows us to train an instance classifier which can be applied in a real-time scenario.

Using this approach, the final bag label is obtained using a max pooling layer.
That is \\
\[
    \widehat{Y_i} = \{\hat{y}_{in}\} = \{ \max\limits_{j}f_n(x_{ij}) \}
\] 
where $f_n(x_{ij})$ is the predicted probability of class $n$ on instance $x_{ij}$. 

The multi-class MIL loss can then be defined as simply the cross entropy loss summed over all the classes, which is \\
\[
    J_i = -\sum_{n}{(y_{in}\log{\hat{y}_{in}} + (1-y_{in})\log{(1-\hat{y}_{in})})}
\] 

\begin{figure}[htb]
  \centering
  \includegraphics[width=\linewidth]{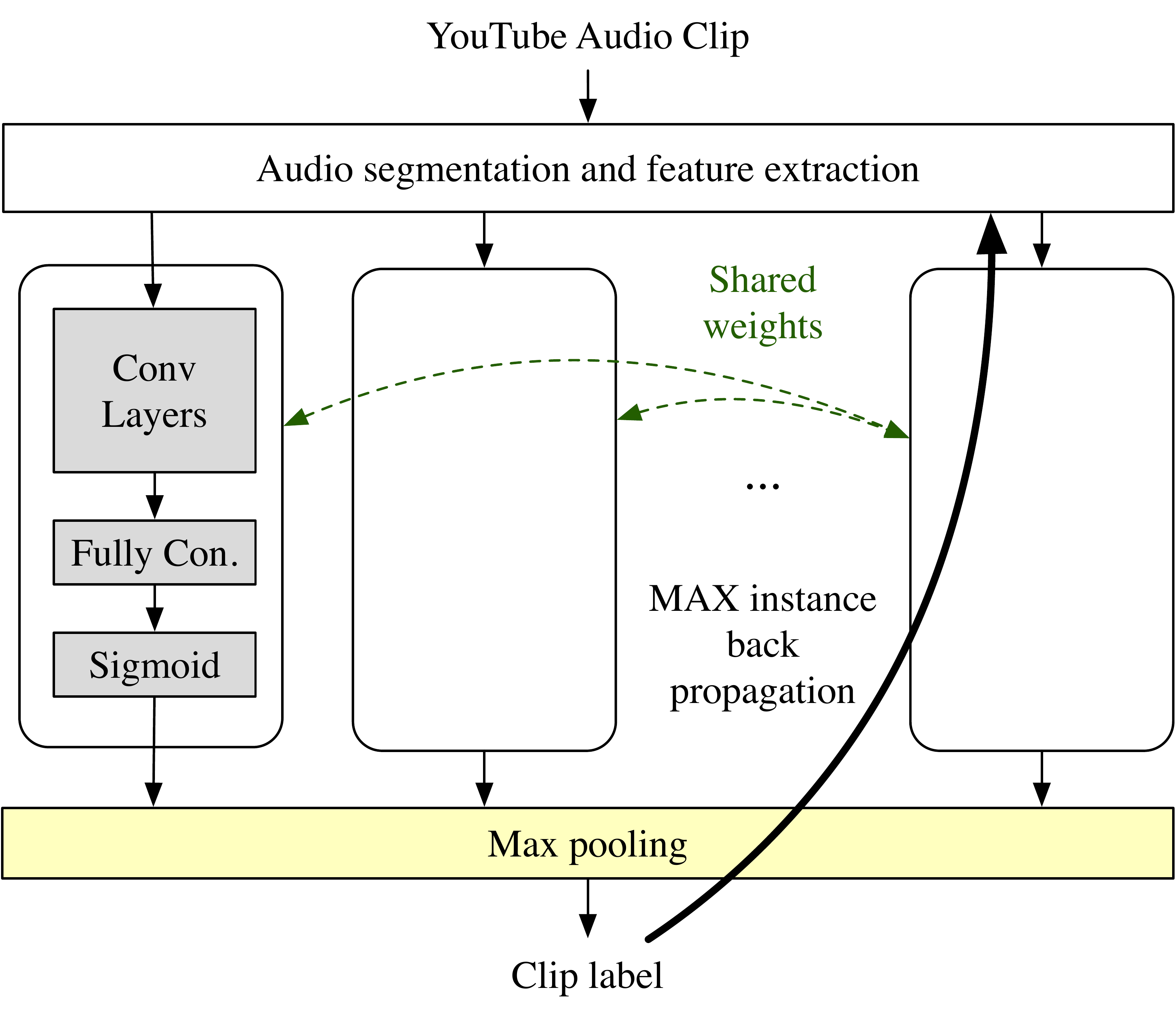}
  \caption{Architecture of MIL using CNN. Back-propagation is performed along the MAX instance for each class.  }
  \label{fig:milcnn}
\end{figure}

In order to address class imbalance we apply a weight to the MIL loss proportional to the inverse frequency of each class.
During back-propagation only the gradient from the maximally scoring instance is calculated and used for updating weights. 
An interesting fact is that as each class has its own max pooling layer, errors originate from different instances between classes. 
Figure \ref{fig:milcnn} shows the architecture of the proposed MIL framework using CNN.

\subsection{MIL using Audio Embeddings}
\label{sec:embeddings}

Our model infers that for a certain class, the highest scoring instances are most important and contribute directly to the corresponding bag label.
The training of the neural network to identify these important instances is similar to an expectation maximization (EM) approach. 
However there are two possible issues which may result from this model.
The first is that as with most EM methods, system performance highly depends on the initialization point.
With a bad initialization point the model chooses the wrong instance as being indicative of the class label and optimizes on irrelevant input. 
These types of errors would be hard to recover from if there is high variation for each individual audio event.
A second issue is that by using a max pooling layer over all instances back-propagation will only propagate through the maximum scoring instance. 
This may result in some instances being ignored for most of the training. 
While this focus on relevant instances only is the central idea of MIL, it greatly reduces robustness to noise which occurs intermittently in the audio. 
We propose that the use of pre-trained audio embeddings can alleviate the above issues.
By using audio embeddings as features we postulate that audio events as well as noise conditions can be better represented which can improve the performance of the MIL framework.

Similar to \cite{Hershey2017CNN-Architectur} we generate audio embeddings by training a CNN to give frame-wise predictions of the clip label.
The input features are 128-bin log-mel spectrograms computed over 1-second segments of audio by short-time Fourier transform.  
We use the clip label as targets for all 1-second segments in the audio clip. 
The outputs from the penultimate layer of the CNN are then extracted and used as input to the MIL framework.
We use the same CNN structure described in the previous section but add an additional fully-connected layer of size 512 to generate the final audio embedding.
Since frame-wise training of the instances results in badly labeled data, the final model selection of the embedding CNN is crucial in generating meaningful embeddings. 
We use the maximum of frame-wise predictions as the predicted clip label and select the CNN model with the best performance at the clip-level using held-out validation data.

\begin{figure}[thb]
  \centering
  \includegraphics[width=\linewidth]{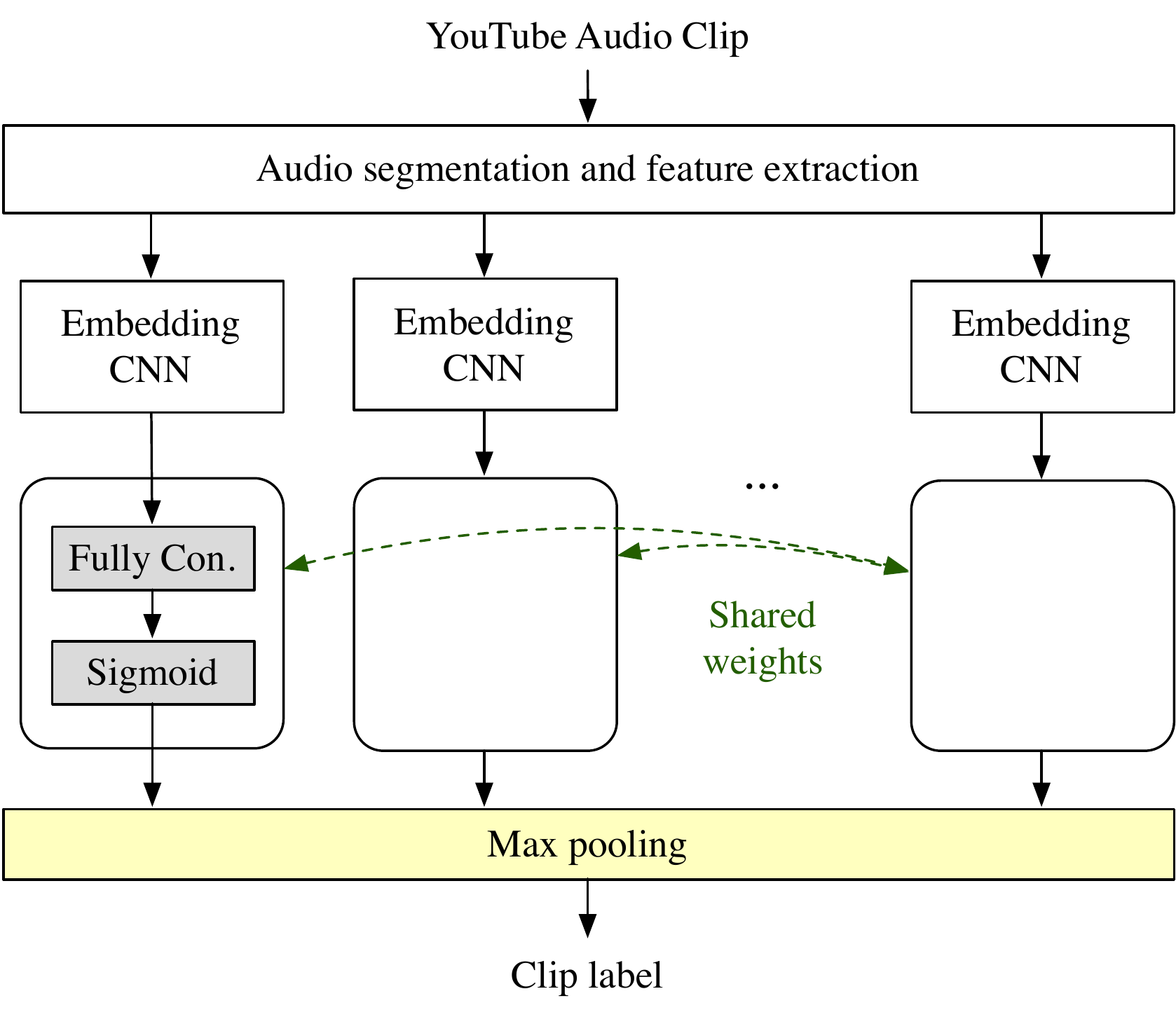}
  \caption{Architecture of MIL using audio embeddings.}
  \label{fig:mildnn}
\end{figure}

The final MIL system is similar in architecture to the MIL-CNN but uses audio embeddings as features for each instance.
The convolutional layers are replaced with fully-connected layers as we no longer deal with images.
The best performing system has four hidden layers using a ReLU activation function with layer sizes of 512, 512, 256 and 128.
The final architecture of the MIL framework is shown in Figure \ref{fig:mildnn}.

\section{Dataset \& Challenges}
\label{sec:format}
\subsection{Dataset}
We evaluated our models using a subset of Google's AudioSet \cite{Gemmeke2017}.
AudioSet is an extensive collection of 10-second YouTube clips annotated over a large number of audio events.
This dataset contains 632 audio event classes and over 2 million sound clips, however as a proof of concept we refer to a subset released by the DCASE 2017 challenge \cite{Badlani2017}.

The challenge subset contains 17 audio event classes divided into two categories : \textit{Warning} and \textit{Vehicle} sounds.
These audio events are highly focused on transportation scenarios and is primed towards evaluating AED systems for self-driving cars, smart cities and related areas.
The subset contains 51,172 samples which is around 142 hours of audio.
The class names and number of samples per class are shown in Table \ref{tab:labels}.

\begin{table}[bth]
    \small
    \centering
    \begin{tabular}{c | c | c | c }
        \textbf{Class Name}    & \textbf{Samp \#} & 
        \textbf{Class Name}    & \textbf{Samp \#}\\\hline
        \textit{Warning Sounds } & & \textit{Vehicle Sounds} & \\\hline
        Car alarm & 273 & Skateboard & 1,617 \\
        Reversing beeps & 337 & Bicycle & 2,020 \\
        Air/Truck horn & 407 & Train & 2,301\\
        Train horn & 441 & Motorcycle & 3,291\\
        Ambulance siren & 624 & Car passing by & 3,724 \\ 
        Screaming & 744 & Bus & 3,745 \\
        Civil defense siren & 1,506  & Truck & 7,090\\  
        Police siren & 2,399  & Car & 25,744 \\
        Fire engine siren & 2,399 & &  \\ 
        
    \end{tabular}
    \vspace{5px}
    \caption{Class labels and number of samples per class.}
    \label{tab:labels}
\end{table}
\subsection{Challenges of the Dataset}
The main challenge of the dataset is the noisiness of YouTube data. 
As clips are user submitted and mostly recorded using consumer devices in real life environments, audio events are often far-field and corrupted with a variety of noise, including human speech, music, wind noise, etc. 
Another challenge is the variability of audio events.
Even within class, the characteristic of an audio event can vary drastically.
An example of this is the use of different types of sirens by different regions which would make it hard to differentiate between \textit{ambulance} and \textit{fire truck sirens}.
In short, it is possible that each label type encompasses all possible global variations of that category.

Finally, the number of samples per class is also highly imbalanced in the subset dataset.
The imbalance ratio of the least occurring to most occurring class is 1:94. 
While this issue can be alleviated through machine learning techniques, the inherent shortage of information in minority classes may result in bad generalization of those classes.


\section{Experimental Setup and Results}
\label{sec:results}
In all experiments we used cross entropy as the loss function and the Adam optimizer \cite{kingma2014adam} to perform weight updates. 
To handle class imbalance the loss function was weighted inversely proportional to the number of samples for each class. 
For model selection of the embedding CNN we adopted a clip-level validation scheme.
The posterior class probabilities were averaged over all instances in a clip and the model with the best clip tagging accuracy was selected to generate audio embeddings.

We compared our MIL framework to an MLP baseline from the DCASE challenge \cite{Badlani2017}.
The best F1-score achieved by our MIL system using a CNN architecture on a two-fold cross-validation setup was 22.4\%.
Using audio embeddings as features and only a DNN as classifier the performance improved to 31.4\% which is 20.5\% absolute improvement from the DCASE baseline. 
We compared to an MIL framework where the DNN classifier is replaced with a 3-layer Bi-LSTM RNN and found that results were comparable to DNNs. 
We also applied late-fusion to models with different hyper-parameters using a weighted majority voting scheme which improved the F1-score further to 35.3\%.
The weights of the voting scheme were based on model validation accuracy.  
Finally, we show that the performance of our MIL framework improves to 46.5\% using embeddings from AudioSet.
These embeddings are part of AudioSet and trained with a CNN architecture from \cite{Hershey2017CNN-Architectur} using the YouTube-8M dataset \cite{abu-el-haija2016_youtube-8m}.
Table \ref{tab:milresults} shows the performance and parameter number of the different models.

\begin{figure}[htb]
  \centering
  \includegraphics[width=\linewidth]{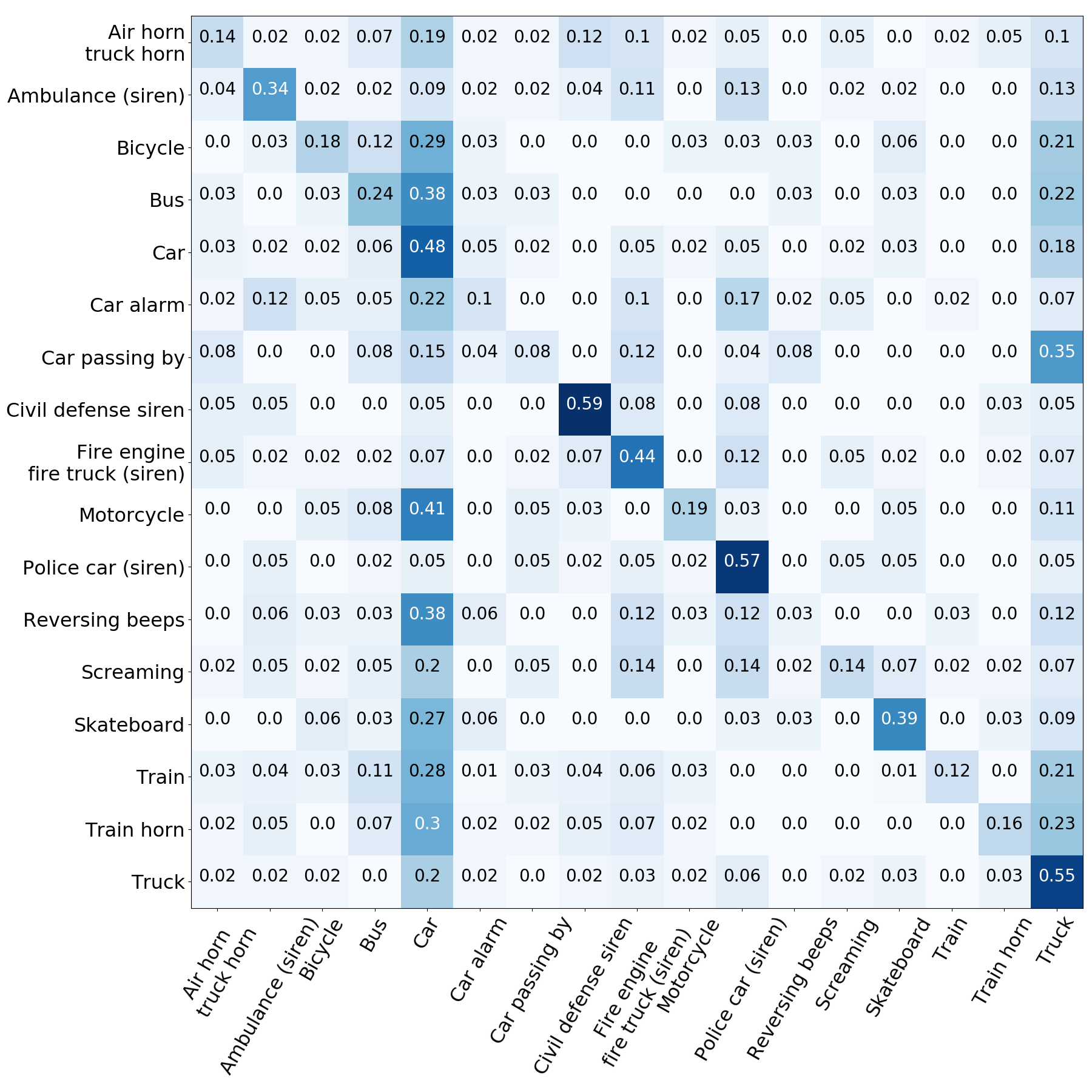}
  \vspace{-20px}
  \caption{Confusion matrix for the proposed MIL system.}
  \label{fig:milcm}
\end{figure}

The confusion matrix for the proposed MIL system is shown in Figure \ref{fig:milcm}. 
Although there is high confusability in the \textit{Car} class, which may be due to the imbalance of labels, the system is still able to distinguish between classes with relative accuracy.

\begin{table}[tbh]
    \small
    \centering
    \begin{tabular}{c| c | c | c | c  }
        \textbf{Model}     &  \textbf{Prec.} & \textbf{Rec.} & \textbf{F1} & \textbf{Param \#} \\\hline 
        \textit{Development set}\\\hline
        Baseline \cite{Badlani2017} & 7.9 & 17.6 & 10.9 & 13K \\
        MIL-CNN  & 19.6 & 26.1 & 22.4 & 29M\\
        MIL-RNN-Embed  & 23.7 & 38.1 & 29.2 & 6.5M \\
        MIL-DNN-Embed  & 25.4 & 41.3 & 31.4 & \textbf{700K}\\
        Ensemble & 28.6 & 46.0 & 35.3 & - \\
        MIL-DNN-AudioSet & 	41.9	& 52.2 & 46.5 & \textbf{700K} \\
        \\
        \textit{Evaluation set}\\\hline
        Baseline \cite{Badlani2017} &  15.0 & 23.1 & 18.2 & 13K \\
        Ensemble & 	31.6	& 39.7 & 35.2 & - \\                         
    \end{tabular}
    \vspace{5px}
    \caption{Comparisons of precision, recall, F1-score (\%), and number of parameters for the various models.}
    \label{tab:milresults}
\end{table}

\section{Discussion}
\label{sec:discussion}
While our framework is not state-of-the-art \cite{xu2017_large-scale-wea}, which achieves a single-model F1-score of 54.2\%, a more fair comparison would be with models without recurrent layers, such as \cite{Chou2017FrameCNN}, which has an F1-score of 49.0\%. 
Even so, a direct comparison is of limited value as our proposed method mainly aims to address two major issues in deploying AED to real-life scenarios: model complexity and real-time operation.  
Our proposed method reduces model complexity by removing the need of recurrent layers and is suitable for applications where computational resources are limited. 
Under similar performance conditions the MIL system using DNN reduces the number of parameters by a factor of almost 10 compared to a 3-layer Bi-LSTM RNN.
In terms of evaluation runtime, the DNN model is also up to 5 times faster than RNNs.
The DNN model is able to handle 2,500 samples per second compared to 500 samples with RNN using an NVIDIA GTX-1080 GPU.

In addition, by using independent instance classifiers our system is able to run in real-time and give running predictions of audio events. 
This property is crucial when applying AED in smart cars as events such as sirens and horns have to be detected as soon as they occur.
With recurrent networks or even CNNs requiring full length inputs this mode of operation would not be possible. 

Finally, as shown by the gain in performance through the use of AudioSet embeddings, the MIL system can easily be improved through transfer learning of other sound events. 
An interesting observation from our experiments is that joint optimization of the pre-trained embedding CNN with the MIL loss did not improve performance much above random initialization.
This shows that audio embeddings already contain rich acoustic information and can be trained in a task-independent manner.
The separation of embedding and classifier training means that we can take advantage of additional labels in large-scale weakly-supervised data and learn embeddings independently. 
However, we also observed that selection of the embedding model is pivotal in the final system performance and not all embeddings are as useful. 


\section{Conclusions}
\label{sec:conclusion}

In this work we proposed a small-footprint multiple instance learning framework using deep neural networks for audio event detection which can be trained using large-scale weakly-supervised data.
We showed that by using pre-trained audio embeddings we can achieve good performance with a simple DNN model in an MIL framework. 
Audio embeddings were extracted from a CNN trained to give frame-wise predictions for the weakly labeled data. 
While the performance of this CNN is poor, the embeddings generated by this model can be used as features to drastically improve the performance of an MIL framework. 
Further improvements were achieved by using embeddings from AudioSet which were trained with more data and additional labels. 
We postulate that audio embeddings map data into an acoustically meaningful high-dimensional space which is more indicative of audio events. 
Using these embeddings we can achieve a good trade-off between model size and performance.

In future work, we hope to apply our model to the entire AudioSet for a truly large-scale weakly-supervised MIL framework.
With the introduction of additional data as well as class labels we expect the audio embeddings to contain richer representations which can further improve performance of AED in smart cars.

\bibliographystyle{IEEEtran}

\bibliography{main}

\end{document}